# New method to characterize a machining system: application in turning

Claudiu F. Bisu[1,2], Jean-Yves K'nevez[2], Philippe Darnis[3], Raynald Laheurte[2,3], Alain Gérard[2]

[1] University Politehnica from Bucharest, 313 Splaiul Independentei, 060042 Bucharest Roumanie (UE) email: cfbisu@gmail.com
[2] Université de Bordeaux, 351 cours de la Libération, 33405 Talence-cedex France (UE) email: alain.gerard@u-bordeaux1.fr
[3] Université de Bordeaux - IUT EA 496, 15 rue Naudet, 33175 Gradignan Cedex France (UE)

**Abstract**
Many studies simulates the machining process by using a single degree of freedom spring-mass sytem to model the tool stiffness, or the workpiece stiffness, or the unit tool-workpiece stiffness in modelings 2D. Others impose the tool action, or use more or less complex modelings of the efforts applied by the tool taking account the tool geometry. Thus, all these models remain two-dimensional or sometimes partially three-dimensional. This paper aims at developing an experimental method allowing to determine accurately the real three-dimensional behaviour of a machining system (machine tool, cutting tool, tool-holder and associated system of force metrology six-component dynamometer).

In the work-space model of machining, a new experimental procedure is implemented to determine the machining system elastic behaviour. An experimental study of machining system is presented. We propose a machining system static characterization. A decomposition in two distinct blocks of the system "Workpiece-Tool-Machine" is realized. The block Tool and the block Workpiece are studied and characterized separately by matrix stiffness and displacement (three translations and three rotations). The Castigliano's theory allows us to calculate the total stiffness matrix and the total displacement matrix.

A stiffness center point and a plan of tool tip static displacement are presented in agreement with the turning machining dynamic model and especially during the self induced vibration. These results are necessary to have a good three-dimensional machining system dynamic characterization (presented in a next paper).

**Keywords**: experimental model, displacement plan, self-excited vibrations, turning}

## *Nomenclature*

| | |
|---|---|
| a | Distance between displacement transducer |
| **BT** | Block Tool |
| **BW** | Block Workpiece |
| [$C$] | Damping matrix |
| [$C_O$] | Compliance matrix |
| [$C_i$] | Displacement transducer (i =1 to 6) |
| **CR**$_{BT}$ | Block Tool **BT** stiffness center |
| $D_1$ | Holding fixture diameter (mm) |
| $D_2$ | Workpiece diameter (mm) |
| {D} | Small displacements torsor |
| $D_{ij}$ | Straight line corresponding of the displacement direction of the point $P_{i,j}$ (i = x, y, z) and (j=1, 2, 3) |
| $d_{ij}$ | Points displacements vectors $P_{i,j}$ (i = x, y, z) and (j=1, 2, 3) |
| $D_x$ | Distance between the line $D_{ij}$ |
| E | Young modulus (N/mm²) |
| $e_x$ , $f_x$ | Scale factors |
| $F_i$ | Force vectors applied to obtain **BT** stiffness center (i = x, y, z) |
| I | Inertial moment |
| [$K$] | Stiffness matrix (N/m) |
| [$K_C$] | Stiffness matrix of rotation (Nm/rad) |
| [$K_F$] | Stiffness matrix of displacement (N/m) |
| [$K_{F,BT}$] | Stiffness matrix of **BT** displacement (N/m) |
| [$K_{F,BW}$] | Stiffness matrix of **BW** displacement (N/m) |
| [$K_{F,WAM}$] | Stiffness matrix of machining system displacement (N/m) |
| [$K_{errors\%}$] | Errors matrix for the matrix [$K$] |
| [$K_{CF}$] | Stiffness matrix of rotation / displacement (Nm/m) |
| [$K_{FC}$] | Stiffness matrix of displacement /rotation (N/rad) |
| $L_1$ | Holding fixture length (mm) |
| $L_2$ | Length workpiece (mm) |
| $M_i$ | Point intersection between straight lines ($D_{ij}$)(i = x, y, z) and (j=1, 2, 3) |
| m | Displacement measured at the charge point |
| [$M$] | Mass matrix |
| $\vec{n_i}$ | Plan normal $P_i$ |
| O | Tool tip point |
| $O_C$ | Cub center |
| P | Force (N) |
| $P_i$ | Plan including the point $M_i$ |
| ***P**_{BT}* | Displacement plan considering tool point |
| $P_{ij}$ | Charge points (i = x, y, z) and (j=1, 2, 3) |
| {T} | Mechanical actions torsor |
| [V] | Matrix eigenvector [$K_{F,BT}$] |
| $v_1$ | Matrix eigenvalue [$K_{F,BT}$] |
| **WTM** | Workpiece-Tool-Machine |
| x (z) | Cross (feed) direction |
| y | Cutting axis |
| $\delta$ | Displacement (mm) |
| $\varepsilon_i$ | Displacement along i (i=1,2,3) |
| $\theta$ | Measured angle at the force point |
| $\theta_i$ | Angular deviation of "Co-planarity" between lines $D_{ij}$ (i = x, y, z; and j=1, 2, 3) |
| $\mu_i$ | Minimal distance between straight lines $D_{ij}$ (i = x, y, z; and j=1, 2, 3) |
| $\rho_i$ | Rotation along i (i=x, y, z) |

## *Introduction*

Metal cutting is one of the most important manufacturing process. The most common cutting processes are turning, milling, drilling and grinding.

During the cutting process of different materials, a whole of physic-chemical and dynamic phenomena are involved. Elasto-plastic strains, friction and thermal phenomena are generated in the contact zone between workpiece, tool and chip. These phenomena are influenced by: physical properties of the material to be machined, tool geometry, cutting and lubrication conditions, and also the machining system dynamic parameters (stiffness, damping). The machine tool vibrations are generated by the interaction between the elastic machining system and the cutting process. The elastic system is composed of: the different parts of the machine tool in movement, the workpiece and the tool. Actions of the machining process are usually forces and moments. These actions also generate relative displacements of elements composing the elastic system. They occur for example between the tool and workpiece, the tool device and bed, etc. These displacements modify the cutting conditions and in the same way the forces. Thus, the knowledge of the machining system elastic behaviour is essential to understand the cutting process [6].

Certain scientists developed a finite element beam model of spinning stepped-shaft workpiece to perform stability analysis using Nyquist criterion [38]} or the traditional stability lobe diagram [16, 25]. This traditional stability analysis technique shows that the chatter instability depends on the structural damping in the system and the spindle speed. Chen and Tsao presented a dynamic model of cutting tool with [11] and without tailstock supported workpiece using beam theory [10]. Here, the effects of workpiece parameters are studied on the dynamic stability of turning process by treating the workpiece as a continuous system. Carrino et al., [8] present a model that takes into account both the workpiece deflection and the cutting force between tool and workpiece. The three components of the cutting force are function of the cutting geometry. The effect of the workpiece-tool-machine deflections is a shift of the workpiece cross-section and a moving back of the tool holder in the radial and the tangential direction (2D model).

In these processes, the cutting forces measurement has important and tremendous applications within industry and research alike. The cutting forces estimation allows to supervise tool wear evolution [36], establishes material machinabilities, optimizes cutting parameters, predicts machined workpiece surface quality and study phenomena such as chip formation or vibrations appearance. Sekar and Yang propose a compliant two degree of freedom dynamic cutting force model by considering the relative motion of workpiece with cutting tool. Tool and workpiece are modelled as two separate single degree of freedom spring-mass-damper systems [34].

In the literature, there are many studies concerning the cutting force measurement. Many dynamometers for this purpose have been developed [5, 9, 12, 23].

Independently of the machining operation type, methods for cutting force measurements can be divided into two general categories. The first corresponds to the category that uses the current or the voltage signals emitted by the tool machine drive motor or control systems [27]. The second uses transducers mounted on the tool or the workpiece assemblies [20, 37, 40].

The cutting forces developed in machining operations may be estimated indirectly by obtaining the power consumed or directly from metal cutting dynamometers; mechanical, hydraulic, pneumatic or several types of electro-mechanical dynamometers.

Knowing the cutting forces is essential to machine tool builders in calculating power requirements and frame rigidity. Cutting forces acting on the tool must be measured at the design tool that are strong enough to remove chip at the desired quantity from the workpiece and to calculate power of tool driver system. The dynamometer is able to measure three force components: cutting force ($F_c$), feed force ($F_a$) and radial force ($F_p$) but not the torque at the tool tip. Axinte et al., [1] propose a procedure to account for both calibration and process errors in the uncertainty estimation for the specific situation of single cutting force measurements. The influence parameters considered in their work, contribution to the measurement uncertainty, workpiece, tool and machine were not considered. Perez et al., [28] give a mechanistic model for the estimation of cutting forces in micromilling based on specific cutting pressure. The model includes three parameters which allow to control the entry of the cutter in the workpiece. The errors in the radial and the angular position of the cutting edges of the tool is considered. The cutting forces are calculated on the basis of the engaged cut geometry, the undeformed chip thickness distribution along the cutting edges, and the empirical relationships that relate the cutting forces to the undeformed chip geometry. This relation do not take into account the elasticity of the machining system. In the measurement of the cutting forces [35], only elastic deflections of the cutting tool due to the cutting forces were measured by means of the load cells located at suitable position on the cutting tool.

However the dynamometer can measure three perpendicular cutting force components and three torque components simultaneously during turning, and the measured numerical values can be stored in computer by data acquisition system [12]. This dynamometer was designed to measure up to 5,000 *N* maximum force and 350 *N/m* torque. The system sensitivity is *pm* 4 % in force and *pm* 8 % in torque.

During the cutting process, the cutting tool penetrates into the workpiece due to the relative motion between tool and workpiece and the cutting forces and torques are measured on a measuring plane in the Cartesian coordinate system. The cutting forces have been measured by the dynamometers designed for different working principles as strain gauge based [12]. Thus it is necessary to have a good methodology to measure the workpiece-tool-machine rigidity before measuring forces and torques. This new methodolgy is precisely the purpose of this paper. In section 2 we present the experimental device. After (see section 3) we conceive the workpiece. The workpiece geometry and dimensions retained for these test-tubes were selected using the finite element method coupled to an optimization method, by SAMCEF® software. In the following section 4 a methodology based on the virtual work (three translations and three rotations) is exhibited to study the static aims and to characterize the static equivalent stiffness values in order to identify the three-dimensional elastic behaviour of the machining system. The applied efforts are quantified with a force sensor. The torsor of small displacements, 3 linears and 3 rotations displacement, is measured by six displacement transducer. A stiffness global matrix is deduced with its various results. The sum of the two stiffness matrix displacements block tool and block workpiece determines the stiffness matrix of machining system displacement. By the Castigliano's theorem we determine the angle that characterizes the principal direction of deformation. Before concluding, in section 5 the stiffness center is obtained using the least squares method in the coordinate system based on the tool in O point that is the origin of the coordinate system.

## *General points*

Today, machines tool are very rigid there are less and less geometrically faulty. The vibratory problems are strongly related to the cutting. Ideally, cutting conditions are chosen such that material removal is performed in a stable manner. However, sometimes chatter is unavoidable because of the geometry of the cutting tool and workpiece. In [7] the bulk of the motion during chatter comes from the workpiece since it has a static stiffness that is up to 3.2 times less than the cutting tool. Since it is highly impractical to instrument the workpiece during cutting the end goal is to develop an observer that can transform measurements made at the cutting tool into a prediction about the motion of the workpiece. Dassanayake [13] approaches in the 1D case the dynamic response of the tool holder to the request of the tool which follows a regenerative surface. They consider only tool motions and disregards workpiece vibration. Insperger continues in the 2D case keeping workpiece rigid but he takes into account the flexibility of the tool[18]}. For an operation of milling [33], the deflections of the machine-tool, the toolholder and the toolholder clamping in the spindle, the tool clamping in the toolholder, and the tool itself, were measured experimentally under the effects of known forces. The results of this study show that the stiffness of both the machine and the clamping in the machine-spindle-toolholder-tool system have a similar importance in the displacement of the tool tip (subjected to a cutting force) to the deflection of the tool itself. Thus, it is necessary to identify the elastic behaviour of machine parts [6]. These vibrations are generated and self induced by the cutting process. A conventional lathe with high rigidity is used to study these dynamic phenomena. The Workpiece-Tool-Machine (**WTM**) system is presented on the figure 1 for a turning operation.

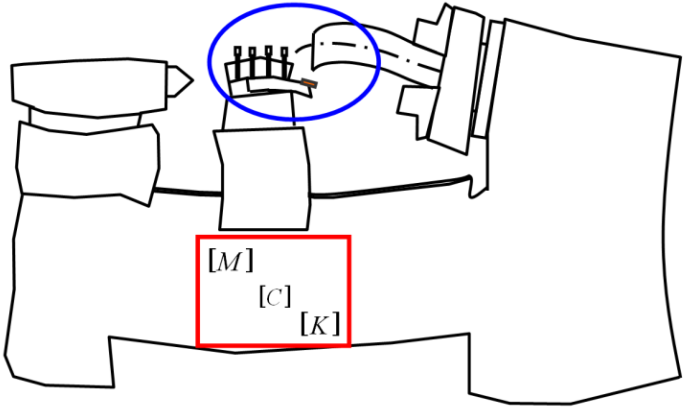

Fig. 1 Workpiece-Tool-Machine considering dynamic cutting process.

The elastic structure of **WTM** system has several degrees of freedom and has many specific vibration modes. The vibrations of each element of the structure are characterized by its natural frequency depending on the Stiffness matrix [K], the Mass matrix [M] and the Damping matrix [C]. In a first time, only the stiffness matrix [K] is studied.

Our experimental approach is based on the matrix development that is presented in [29]. To identify the **WTM** system static behaviour, the machining system is divided into two blocks, the block Tool (**BT**) and the block Workpiece (**BW**) figure 2. These two blocks are related to the turn bed supposed to be infinitely rigid.

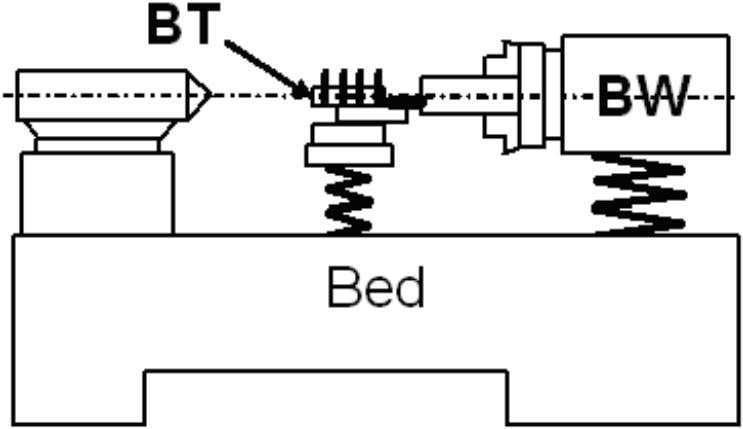

Fig. 2 Presentation of the experimental device.}

## 3 Components of the system WTM

### 3.1 Block Workpiece: BW

As many workers [2, 27, 41], a cylindrical geometry of the workpiece is chosen. The **BW** represents the revolving part of the **WTM** system; it includes the holding fixture, the workpiece and the spindle (figures 3 a, b). To make the whole frame rigid, a very rigid unit (workpiece, holding fixture) is conceived in front of the **WTM** elements (figure 4).

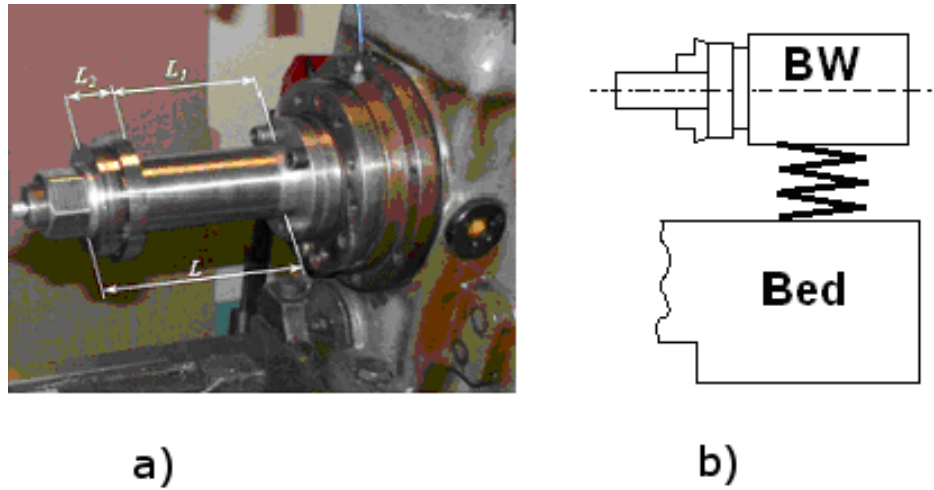

Fig.3 **BW** representation.}

The workpiece geometry and his holding fixture are selected with $D_1 = 60$ mm, $D_2 = 120$ mm and $L_2 = 30$ mm (cf. figure 4). These dimensions retained for these test-tubes were selected using the finite element method coupled to an optimization method by SAMCEF® software. It is necessary to determine the holding fixture length $L_1$ to obtain a significant stiffness in flexion. The objective is to move away the first **BW** vibration mode of the lathe fundamental natural vibration mode (see [3]).

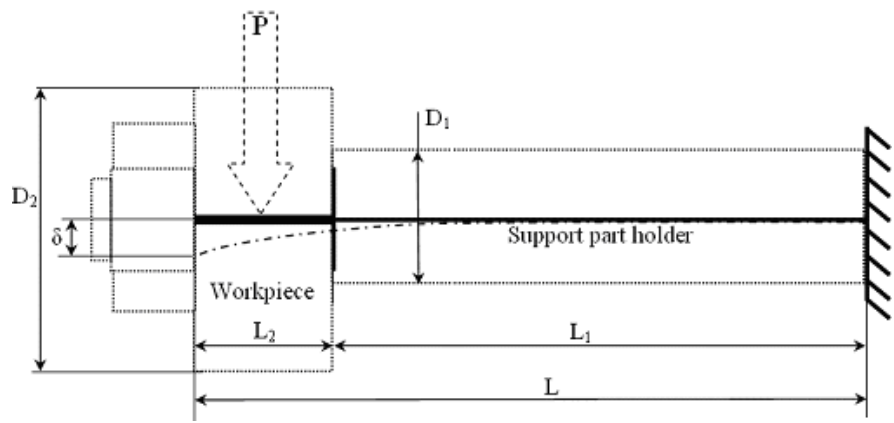

Fig. 4 Geometry of holding fixture / workpiece.}

As [33] and others the stiffness is calculated on the basis of the displacement δ for a given force P :

$$\delta = \frac{P \times L^3}{3E \times I} \quad (1)$$

with inertial moment :

$$I = \frac{\pi \times D_1^4}{64} \quad (2)$$

The figure 5 represents the displacements and stiffness values relating to the length of holding fixture / workpiece, for a force P = 1,000 N, a Young modulus E = 21.10$^{-5}$ N/mm$^2$ and a holding fixture diameter $D_1$ = 60 mm.

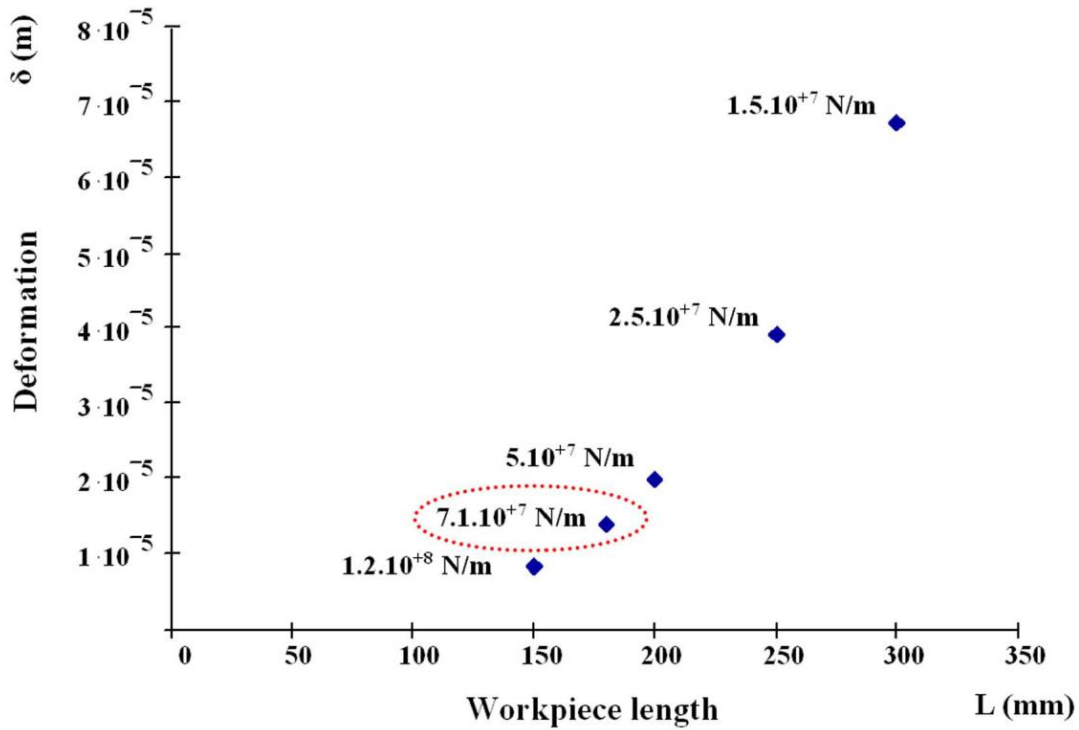

Fig. 5 Displacements according to the holding fixture length.

A holding fixture length : $L_1$ = 180 mm, for a stiffness in flexion of 7.10$^{-7}$ N/m, is reminded. This value is including in the higher part of the interval of the acceptable rigidity values for conventional lathe (cf. figure 6), [19, 21, 22].

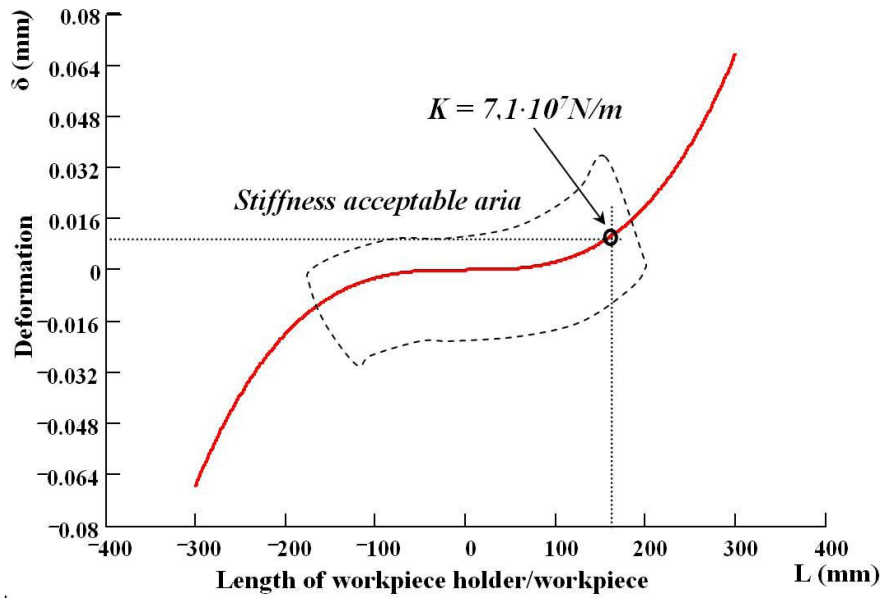

Fig. 6 Acceptable aria representation of the workpiece deformation.}

## 3.2 Block Tool: BT

In this case, the **BT** part includes the tool, the tool-holder, the dynamometer, the fixing plate on the cross slide (figure 7a). The six-component dynamometer [12] is fixed between the cross slide and the tool-holder. This is necessary thereafter to measure the cutting mechanical actions. The stiffness of **BT** is evaluated into the next section.

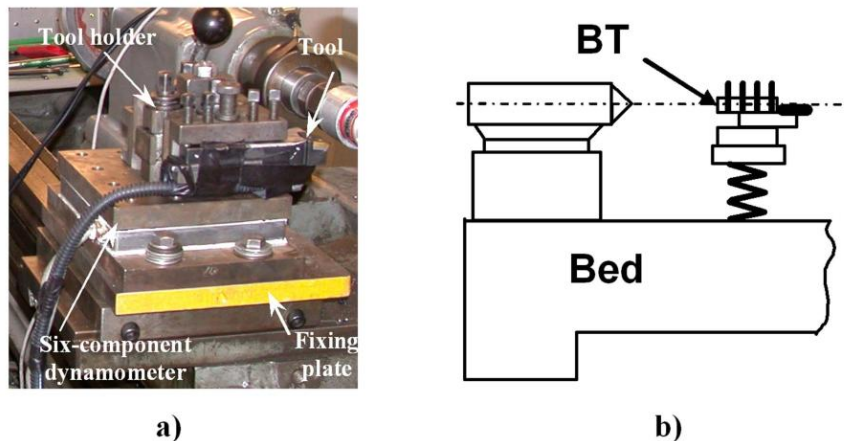

Fig. 7 Block tool **BT** representation.

## *4. Static characterization of machining system*

The static study aims at characterizing the static equivalent stiffness values in order to identify the three dimensional elastic behaviour of the machining system. Generally, the static tests consist in charging by known efforts the two blocks and measuring only the associated displacements components [17, 39]. Here, also the static tests consist in loading by known efforts the two blocks but measuring the small displacements torsor (i.e. three linear displacements, and three rotations). The applied efforts are quantified with a force sensor. The small displacements torsor is measured by six displacement transducers. A stiffness global matrix is deduced with its various results. It is a real 3D pattern. For instance, Carrino et al., present a model that takes into account both the workpiece deflection and the cutting force between tool and workpiece. The three components of the cutting force are a depend on the cutting geometry. The effect of the workpiece-tool-machine deflections is a shift of the workpiece cross-section and a moving back of the tool holder in the radial and the tangential direction (2D model) [8].

## 4.1 Stiffness matrix

The experimental approach is based on the matrix development presented in [30]. The deformation of a structure element is represented by displacements of nodes determining this element. The "associated forces" correspond to displacements which act as these nodes.

The transformation matrix which connects generalized displacements of an element to the "associated forces" is the rigidity matrix or the stiffness matrix of the element. In the same way the matrix which connects generalized displacements of the structure to the applied generalized discrete forces is the stiffness matrix of the structure simply named as "stiffness matrix" : $[K]$.

The relation between forces and displacements is given by [29]:

$$\{T\} = [K] \times \{D\} \qquad (3)$$

where $\{T\}$ represents the mechanical action torsor, $[K]$ the stiffness matrix and $\{D\}$ the small displacements torsor.

The general form of the square (6 x 6) stiffness matrix $[K]$ is:

$$[K]_{A,\,xyz} = \begin{bmatrix} K_{FC} & K_F \\ K_C & K_{CF} \end{bmatrix}_{A,xyz} \qquad (4)$$

where $[K_F]$, $[K_C]$, $[K_{CF}]$ and $[K_{FC}]$ are respectively square (3 x 3) displacement matrix, rotation and rotations / displacement and displacement / rotation expressed at the point A in x, y, z machine axes.

## 4.2 Experimental determination of the stiffness matrix

The matrix elements of the small displacement torsor are identified thanks to the experimental device presented in the figure 8. The considered system is a cube. Displacements are measured by six displacement transducers. Two displacement transducers are positioned symmetrically on each of the 3 directions. The force is applied to each x, y, z direction in two different levels by a screw-swivel system controlled by a force sensor. Each loading point coordinates are known starting from the cube center $O_c$. This allows for each applied force to determine the moment and thus the complete torsor of mechanical actions $\{T\}$.

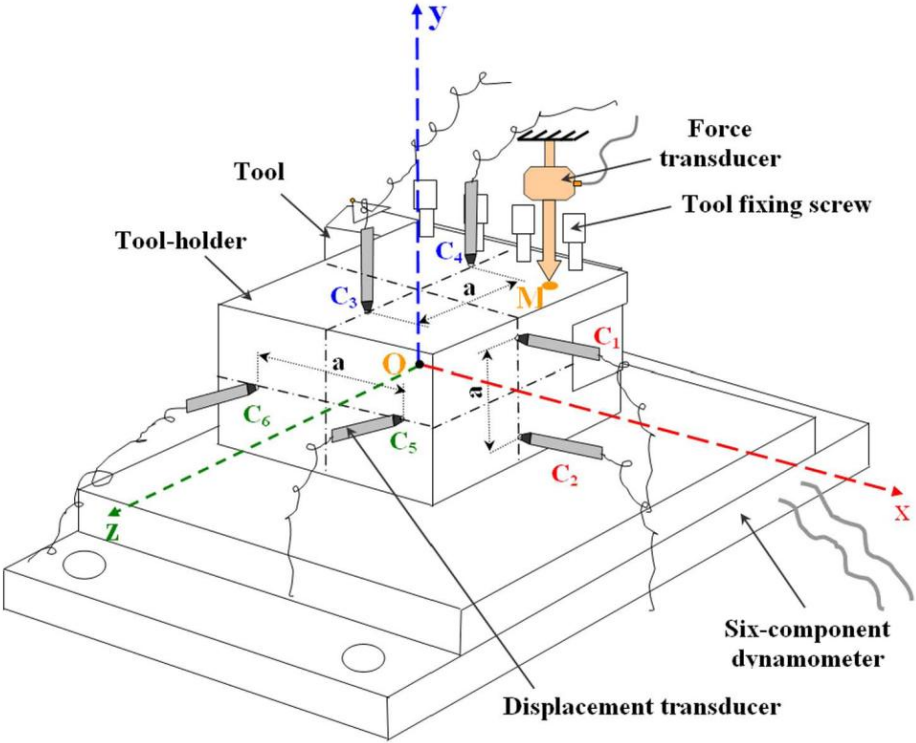

Fig. 8 Experimental device for the static characterization.

Induced displacements are solid body displacements and it is noted that rotations are low (> $10^{-5}$ rad) but exist. The existence of these rotations is important and in agreement with the torque via the virtual work theory. The measure principle is presented in the figure 9 and is used to determine the components of the small displacements torsor $\{D\}$ which is composed by the three rotations $\rho_x$, $\rho_y$, $\rho_z$ and the three displacements $\varepsilon_x$, $\varepsilon_y$, $\varepsilon_z$.

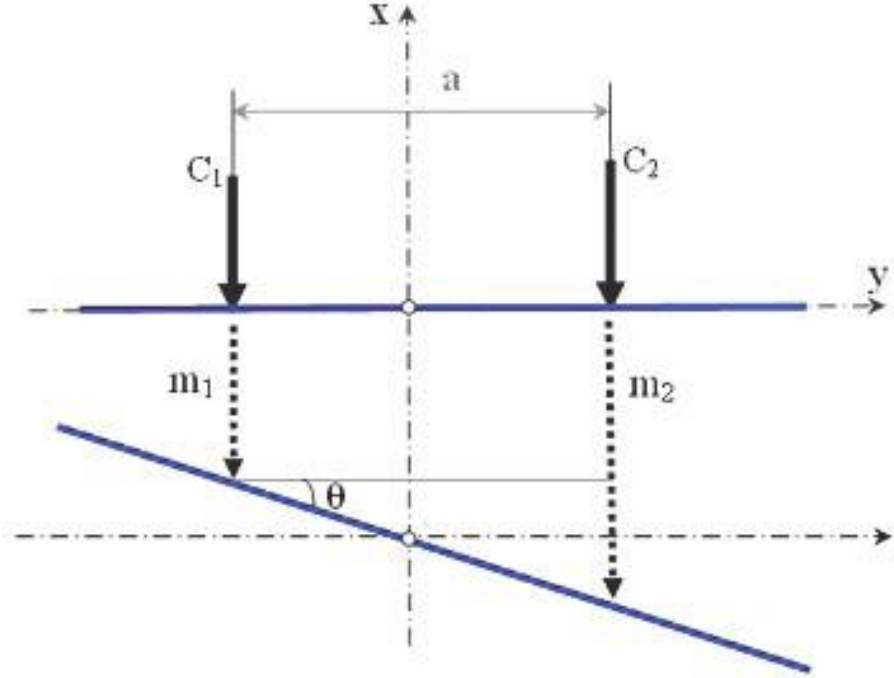

Fig. 9 Position of Displacement transducer.

Thus, the displacements $m$ and rotations $\theta$ are determined, for each loading direction, by using the relations:

$$m = \frac{m_1 + m_2}{2}, \quad \tan\theta \approx \theta = \frac{m_2 - m_1}{a} \tag{5}$$

From these relations and considering the six measurements points it results:

$$\begin{bmatrix} \rho_x \\ \varepsilon_y \\ \varepsilon_z \end{bmatrix} = \begin{bmatrix} 0 & 0 & \frac{-1}{a} & \frac{1}{a} & 0 & 0 \\ 0 & 0 & 0 & 0 & \frac{-1}{a} & \frac{1}{a} \\ \frac{-1}{a} & \frac{1}{a} & 0 & 0 & 0 & 0 \\ \frac{1}{2} & \frac{1}{2} & 0 & 0 & 0 & 0 \\ 0 & 0 & \frac{1}{2} & \frac{1}{2} & 0 & 0 \\ 0 & 0 & 0 & 0 & \frac{1}{2} & \frac{1}{2} \end{bmatrix} \times \begin{bmatrix} m_1 \\ m_2 \\ m_3 \\ m_4 \\ m_5 \\ m_6 \end{bmatrix} \tag{6}$$

The tests are carried out with specific assemblies which are designed for each direction of measurement. The loading (respectively unloading) is carried out by step of 30 daN (resp. − 30 daN) until (resp. from) level of 200 daN, and this procedure is used for each test following known directions. To check the repeatability and accuracy of identifications all tests and measurements are carried out five times and the average is selected for each point at figure 10.

To exploit measurements as well as possible, the displacements curves are plotted depending on the applied force for each loading direction. A line of least squares is adjusted to determine the displacements components values for a given force. Thus, six torsors of small displacements are identified for six loadings cases. The linear behaviour observed in loading is different from the linear behaviour noted in unloading. This different linearity between loading and unloading is due to the existence of deviations and friction forces in each point surfaces of the assembly. These deviations due to the installation of the parts of the associated assembly and friction force are different in charge and discharge.

When this difference in linear behaviour appears (hysteresis), we use the line (figure 10) which passes by the middle (*C* point) of segment *AB* (charge-discharge). *OC* is the line which the slope, by assumption, corresponds to the "real" stiffness. Segment *AB* represents the double of the friction forces and deviations for the deformation δ, [24].

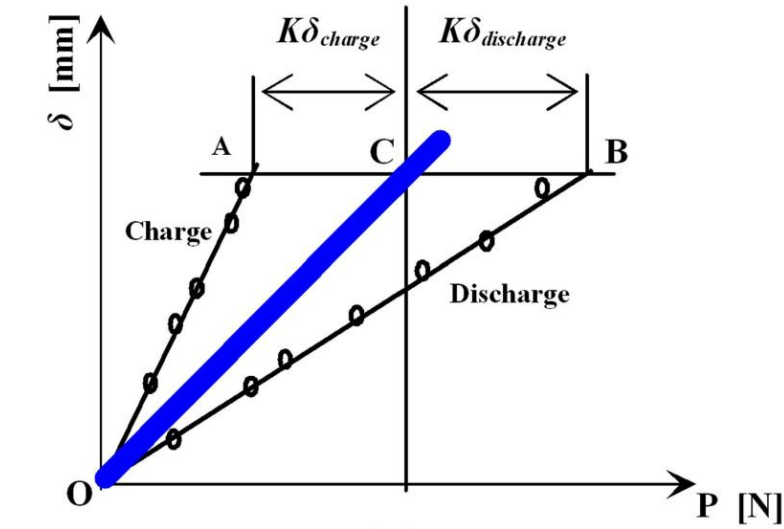

Fig. 10 Diagrammatic representation of linear behaviour in charge and discharge.}

At this stage, the matrix column of the small displacements torsors and of the mechanical action torsors are known. The flexibility matrix $[C_0]$ of the system is deducted by:

$$[C_O] = \{T\}^{-1} \times \{D\} . \qquad (7)$$

The inversion of experimental flexibility matrix gives the global stiffness matrix, $[K]$.

4.2.1 Stiffness matrix of **BT**

The figure 11 presents a loading example in *x* direction of **BT** (see [3]). A similar loading experimentation in y and z direction of **BT** is carried out. For each x, y or z direction measurements are taken at three points and captured by data acquisition card installed in PC. The stored data can be retrieved and used for analysis when required. Using Labview software we obtain the experimental flexibility matrix $[C_0]$. A simple inversion gives the stiffness global matrix $[K_{BT}]$.

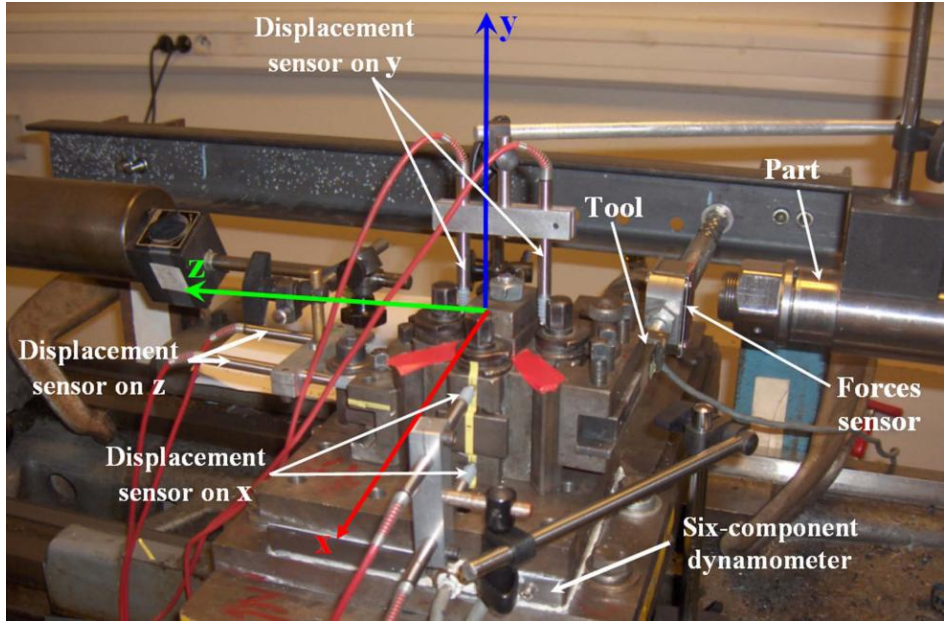

Fig. 11 Example of loading in *x* **BT** direction.

$$[K_{BT}] = \begin{bmatrix} -5\times10^6 & -7.7\times10^6 & 2.9\times10^6 & 6.7\times10^6 & 8.7\times10^5 & -3.4\times10^6 \\ 5\times10^6 & 7.8\times10^6 & -3\times10^6 & -7.5\times10^6 & -3.3\times10^5_v & 1.7\times10^6 \\ 2.7\times10^6 & 4\times10^6 & 5.8\times10^6 & -1.4\times10^6 & -1.7\times10^6 & 1.3\times10^7 \\ -1.8\times10^6 & -2.8\times10^6 & -7\times10^6 & 1.5\times10^6 & -1.6\times10^6 & -1.2\times10^7 \\ 1.7\times10^6 & 2\times10^6 & 4.6\times10^6 & 1.5\times10^6 & -4.7\times10^5 & 5.8\times10^6 \\ 1\times10^6 & 1.4\times10^6 & -7.4\times10^{-5} & -1.9\times10^6 & -2\times10^5 & 1.7\times10^6 \end{bmatrix}_{O,xyz} \quad (8)$$

In addition, on figure 10 we made a simplifying assumption while retaining for each level of deformation given (for example *AB*), the average charge (*C*) correspondent between the value in charge (*A*) and the value discharges (*B*) from them. Consequently, it is advisable to make sure the validity of this assumption. This must be done while estimating, on each level of loading, the error made by using the median value between the charge and the discharge. However, the use of the least squares method allows the evaluation of the error made for each level of loading, i.e. each elements of the matrix $[K_{BT}]$. Thus, a matrix of error can be built. This matrix noted $[K_{errors\%}]$ thus allows to know the error attached to each element of the matrix $[K_{BT}]$. This operation is performed by an errors matrix $[K_{errors\%}]$ given in (9).

$$[K_{errors\%}] = \begin{bmatrix} 0.1 & 0.7 & 3.8 & 2.3 & 5.7 & 1.7 \\ 0.6 & 1.4 & 1.2 & 2.5 & 3.3 & 2.5 \\ 4.3 & 1.2 & 0.05 & 4.4 & 0.1 & 4.7 \\ 0.4 & 0.1 & 0.1 & 0.4 & 1 & 2.2 \\ 0.1 & 0.1 & 0.05 & 0.7 & 0.1 & 0.3 \\ 0.1 & 0.2 & 0.3 & 1.3 & 0.2 & 1.2 \end{bmatrix} \quad (9)$$

It is noted that the error does not exceed 6 % what is largely acceptable.

In addition, for a "perfect" decoupled system [$K$] is diagonal, and the elements are stiffness values in N/m. The matrix [$K$] obtained here is a matrix block. Comparing this matrix to the matrix [$K$]$_{A, xyz}$ (4) we establish between elements the following correspondences:

- the elements of the matrix 3 × 3 in the right higher corner are the elements corresponding to stiffness values of displacements (N/m) noted [$K_F$] in (4),
- the elements of the matrix 3 × 3 in the left lower corner are the elements corresponding to stiffness values of rotations (N/rad) noted [$K_C$] in (4),
- \item the elements of the matrix 3 × 3 in the left higher corner are the elements corresponding to the couplings of "displacements / rotations" noted [$K_{FC}$] in (4),
- \item the elements of the matrix 3 × 3 in the right lower corner are the elements corresponding to the couplings of "rotations / displacements" noted [$K_{CF}$] in (4).

These two last elements ("displacements / rotations" [$K_{FC}$], "rotations / displacements" [$K_{CF}$]) are not taken into account here. Only the stiffness part of displacement (noted [$K_{F, BT}$] below) is necessary considering in our next dynamic model (not presented here).

$$[K_{F, BT}] = \begin{bmatrix} 6.7 \times 10^6 & 8.7 \times 10^5 & 3.4 \times 10^6 \\ -7.5 \times 10^6 & -3.3 \times 10^5 & 1.7 \times 10^6 \\ -1.4 \times 10^6 & 1.7 \times 10^6 & 1.3 \times 10^7 \end{bmatrix}_{O, xyz} \quad (10)$$

4.2.2 Stiffness matrix of **BW**

As the **BW** geometry was simpler, we limited ourselves to measure the stiffness values of displacements by using three displacement transducer following the three main directions. The loading was carried out with the dynamometer (cf. figure 12).

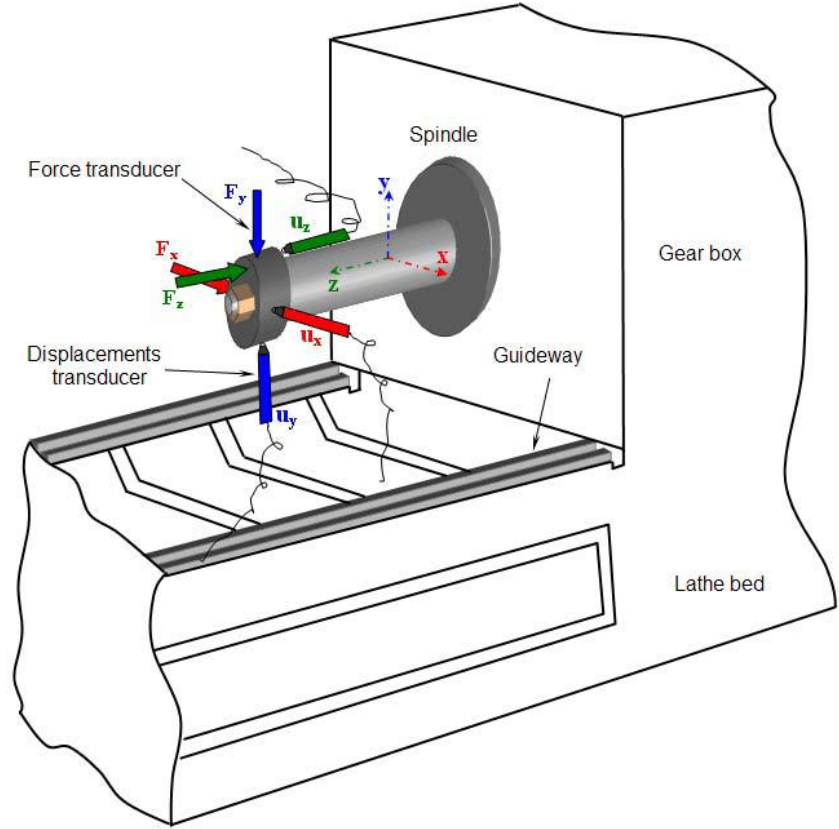

Fig. 12 Experimental device for the **BW** static characterization.

Using this geometry, the **BW** has a very high rigidity according to *z* axis and this value is very small compared with the principal stiffness value. The **BW** behaviour is linear, with a nearly null hysteresis. At the loading and unloading points, the part is not influenced by the friction phenomenon or other various fits generated by the assembled elements, like the spindle or the ball bearings. The stiffness matrix $[K_{F, BW}]$ obtained according to the three directions is:

$$[K_{F, BW}] = \begin{bmatrix} -1.4 \times 10^7 & 0 & 0 \\ 0 & 2 \times 10^7 & 0 \\ 0 & 0 & 2.85 \times 10^8 \end{bmatrix}_{O,xyz} . \tag{11}$$

We may notice that the spindle and its ball bearings decrease the global **BW** rigidity compared with the calculated rigidity in the section 3.1.

### 4.3 Experimental determination of the machining system stiffness matrix

In order to know stiffness values of the machining system in the three directions, the elastic interaction **BT BW** is modelled by static stiffness which are assembled in parallel (figure 13). The chip is the common point between deformation and force. It connects the two stiffnesses (**BT** / **BW**). The sum of the two stiffness matrix of displacements **BT** [$K_{F,BT}$] and **BW** [$K_{F,BW}$] determines the stiffness matrix of machining system displacement [$K_{F,WAM}$].

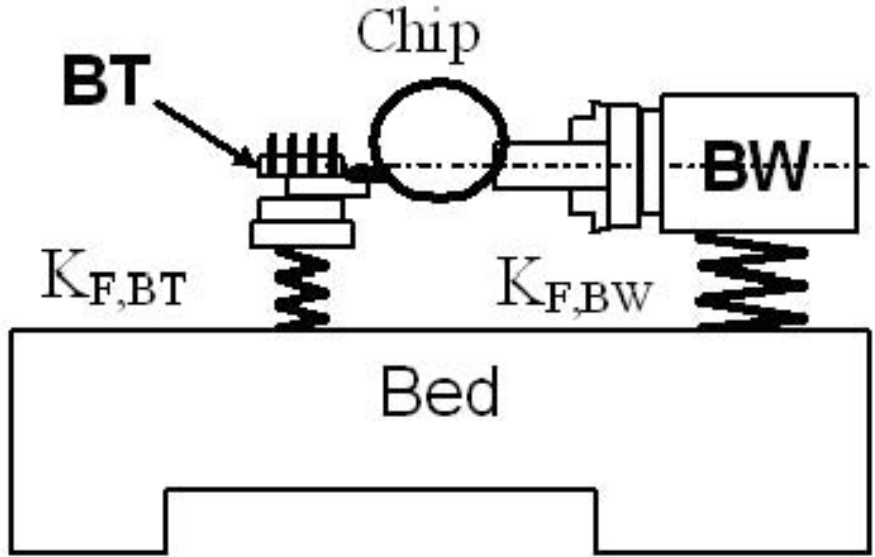

Fig. 13 Static stiffness assembled in parallel.

$$[K_{F,WAM}] = \begin{bmatrix} 2.7 \times 10^7 & 8.7 \times 10^5 & -3.4 \times 10^6 \\ -7.5 \times 10^6 & 2.1 \times 10^7 & 1.7 \times 10^6 \\ -1.4 \times 10^6 & -1.7 \times 10^6 & 2.9 \times 10^8 \end{bmatrix}_{O,xyz} \quad (12)$$

In this form the matrix of the machining system shows that the main diagonal has elements of a higher order than the others. Moreover the diagonalization of the matrix is possible and it comes:

$$[K_{F,\ WAM-d}] = \begin{bmatrix} 2.6 \times 10^7 & 0 & 0 \\ -0 & 2.2 \times 10^7 & 0 \\ 0 & & 2.9 \times 10^8 \end{bmatrix}_{O,xyz} \quad (13)$$

By the Castigliano's theorem we determine the angle $\alpha_K$. This angle characterizes the principal direction of deformation. It takes place in the case of two blocks **BT** and **BW** interaction (figure 14). On the figure 14, $\beta_{K-BT}$ et $\theta_{K-BW}$ correspond respectively to the principal direction of deformation of the Block Tool and Block Workpiece in connexion with the machine axes.

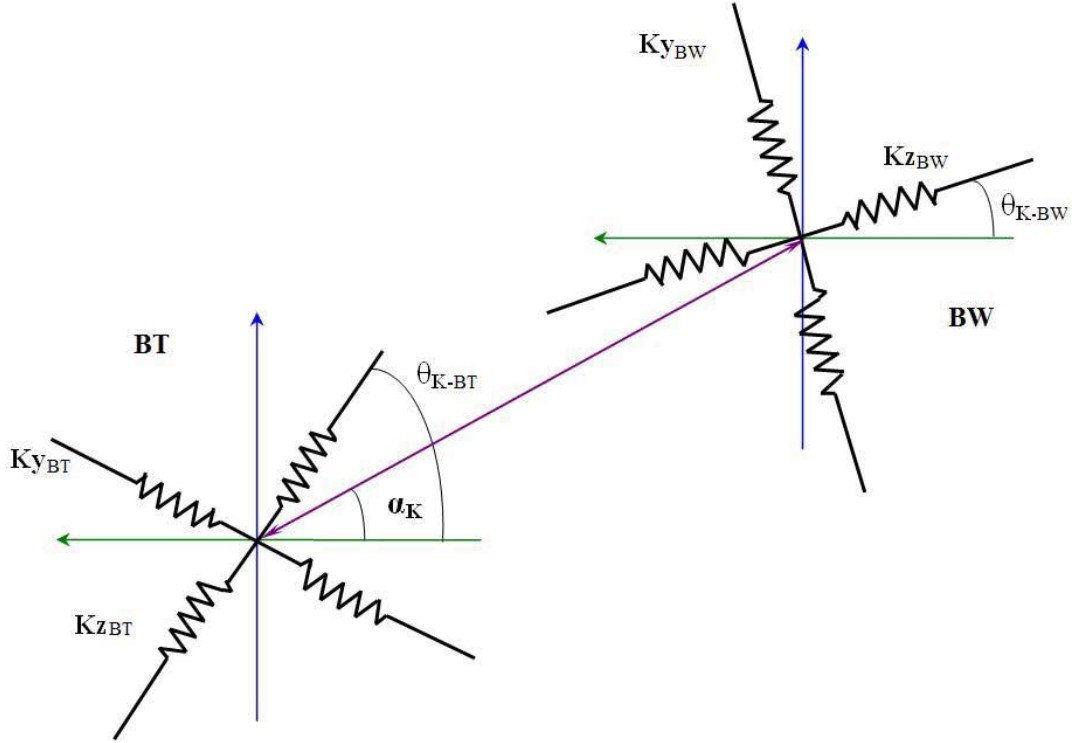

Fig. 14 Diagram of $\alpha_K$ angle determination.

The stiffness values $[K_{BT}]$ and $[K_{BW}]$ of the whole elastic system, among principales axes, are determined by deformation energy minimization, allowing thus the diagonalization of the matrix $[K_{F,BT-d}]$:

$$[K_{F,\,BT-d}] = \begin{bmatrix} 4.1 \times 10^5 & 0 & 0 \\ & 6 \times 10^6 & 0 \\ & & 1.3 \times 10^7 \end{bmatrix}_{O,xyz} \quad (14)$$

In the plane O,x,y, with $\theta_{K-BT} = 52°$ and $\theta_{K-BW} = 0°$ we obtain an angle $\alpha_K = 76°$. In the plane O,y,z with $\theta_{K-BT} = 32°$ and $\beta_{K-BW} = 0°$ we have an angle $\alpha_K = 65°$. On this direction, the maximum deformation of the system is obtained.

## *5 Rotation center*

Deacu [14] and Kudinov [24] show that any machine tool, is characterized by the deformation principal directions. These deformation principal directions are function of the machine structure, its geometrical configurations and cutting parameters used. We can observe either a very stiff behaviour, or very rubber band according to the direction. Here, the tool is regarded as forming integral part of block **BT**. The aim is to determine the stiffness center **CR**$_{BT}$ of elastic system **BT** [26]. This stiffness center corresponds to the rotation center of the block **BT** compared to the bed. Obtaining **CR**$_{BT}$ consists in finding the points of intersection of different perpendicular to displacements.

## 5.1 Experimental step

The stiffness center is obtained in the coordinate system based on the tool in O point that is the origin of the (x, y, z) coordinate system.

The procedure to obtain the stiffness center is detailed in the figure 15; the imposed charge follows all three directions, $F_i$ being known; (i = x, y, z).

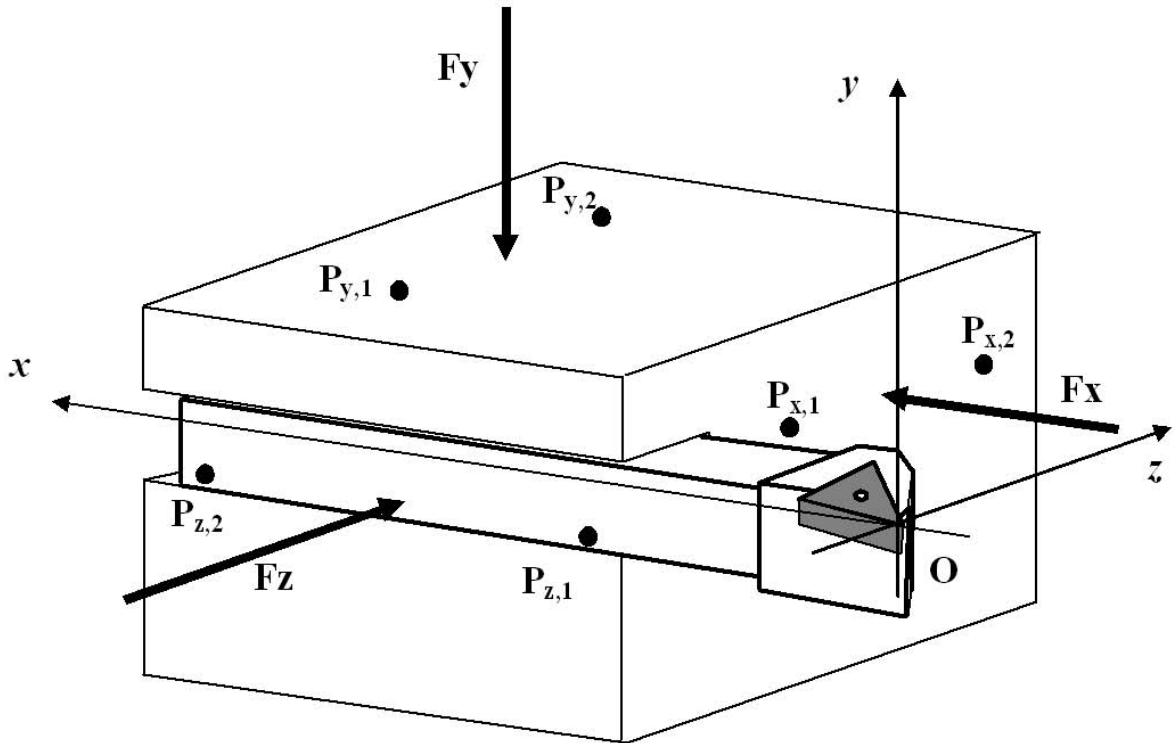

Fig. 15 Experimental procedure to obtain stiffness center **CR**$_{BT}$

The displacement vector measurement under the load $d_{i,j}$ is obtained in two points $P_{i,j}$ in each direction (i = x, y, z) and (j = 1, 2). The direction ($D_{ij}$) is released being straight line containing $d_{i,j}$ vector that include the point $P_{i,j}$.

According to the figure 16, these lines are not exactly coplanar and no secant. It is possible to find for each direction the intersection point, named $M_i$, of the lines $D_{ij}$ using the following relations, for *x* direction:

$$\overrightarrow{OM_x} = \left(\overrightarrow{OP_{x,1}} + \vec{e}_x\right) \times \overrightarrow{Od_{x,1}} \tag{15}$$

Using the method of least squares, we can minimize the distance $d_x$ between the lines ($D_{x,1}$ and $D_{x,2}$, figure 16); $d_x$ is calculated using the expression:

$$d_x = \left| \frac{\left(\overrightarrow{OM_x} - \overrightarrow{OP_{x,2}}\right) \wedge \overrightarrow{d_{x,2}}}{\left\|\overrightarrow{d_{x,2}}\right\|} \right| \tag{16}$$

and the angle $\theta_x$ corresponds of angular deviation of "coplanarity":

$$\theta_x = \arccos\left(\frac{\overrightarrow{P_{x,1}M_x} \times \overrightarrow{d_{x,1}}}{\left\|\overrightarrow{P_{x,1}M_x}\right\| \times \left\|\overrightarrow{d_{x,1}}\right\|}\right) \tag{17}$$

The two lines ($D_{ij}$) are separated by a minimum distance $\mu_i = 1.8$ mm and maximum deviation $\delta_i = 2°$ (see the table 1).

In the next step, three mean plans $P_i$ are defined by approximation of the points $M_i$ and containing the lines $D_{ij}$.

We can draw the normal $\overrightarrow{n_i}$ of each plan $P_i$ that includes the point $M_i$ (figure 17).

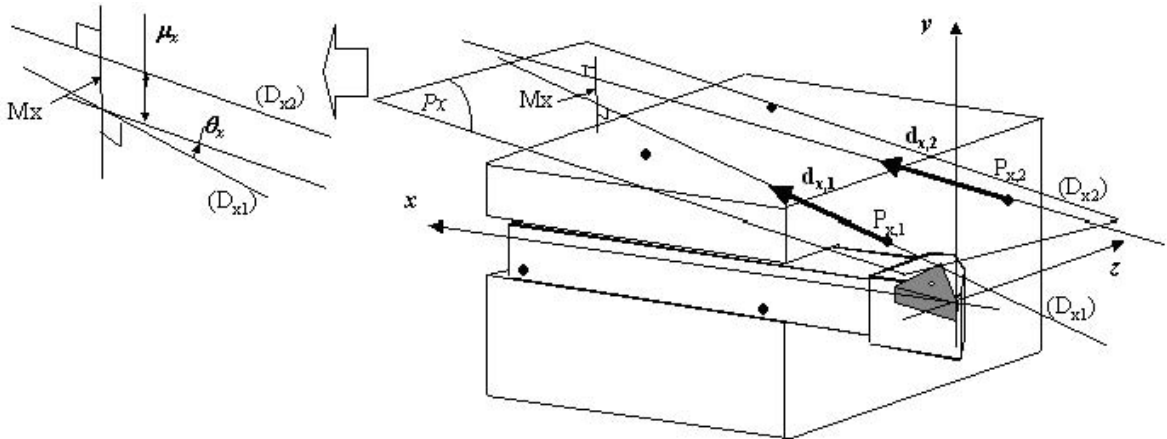

Fig. 16 Identifying the intersection point $M_x$

Writing the loop of geometrical closure under the three directions:

$$\overrightarrow{OCR_{BT}} = \overrightarrow{OJ_i} + f_i \times \overrightarrow{n_i} \tag{18}$$

where: i = x, y, z.

We obtain a linear system with three vectors' equations and three unknown factors, after that is seeking to obtain the intersection point $CR_{BT}$ using least squares minimization method. In reality, the directions of these three normal lines cross (nearly at r = 1.2 mm) in only one point noted **CR**$_{BT}$ (see figure 17) which corresponds with stiffness center.

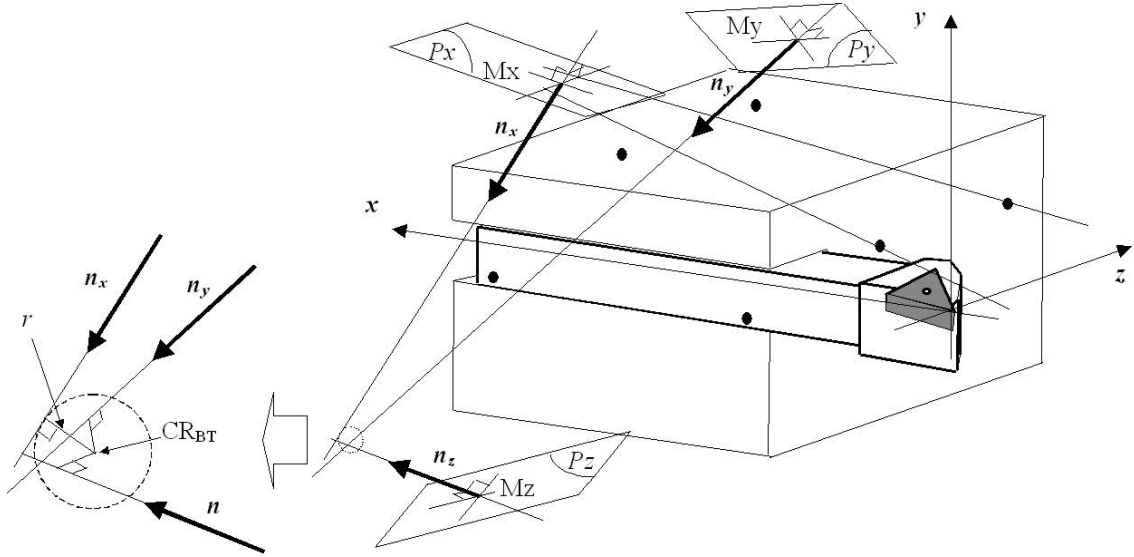

Fig. 17 Determination of the block tool rotation center: **CR**$_{BT}$

| Charge point | Coordinate of point (mm) | Displacement vector $\overrightarrow{d_{ij}}$ (m) | Shift distance $\mu_I$ (m) | Angular shift $\theta_i$ (°) | Intersection point $M_i$ (m) | Coordinate of CR$_{BT}$ (m) |
|---|---|---|---|---|---|---|
| $P_{x,1}$ | 35 | 9.1×10$^{-5}$ | 1.8×10$^{-3}$ | 0.28 | 0.366 | 0.56 |
|  | -20 | 1.7×10$^{-5}$ |  |  | 0.042 |  |
|  | 52 | 3.4×10$^{-5}$ |  |  | 0.178 |  |
| $P_{x,2}$ | 35 | 0.8×10$^{-5}$ | 1.8×10$^{-3}$ | 0.28 | 0.366 | 0.56 |
|  | -20 | 1.5×10$^{-5}$ |  |  | 0.042 |  |
|  | 117 | 1.5×10$^{-5}$ |  |  | 0.178 |  |
| $P_{y,1}$ | 116 | 2×10$^{-5}$ | 1.7×10$^{-3}$ | 1.97 | 0.086 | -0.58 |
|  | 15 | -2×10$^{-5}$ |  |  | 0.045 |  |
|  | 56 | -2×10$^{-5}$ |  |  | 0.081 |  |
| $P_{y,2}$ | 116 | 0.2×10$^{-5}$ | 1.7×10$^{-3}$ | 1.97 | 0.086 | -0.58 |
|  | 15 | -1.3×10$^{-5}$ |  |  | 0.045 |  |
|  | 103 | 9.8×10$^{-6}$ |  |  | 0.081 |  |
| $P_{z,1}$ | 45 | 5.5×10$^{-6}$ | 8.8×10$^{-4}$ | 0.18 | 0.033 | -0.08 |
|  | -20 | 6.5×10$^{-5}$ |  |  | -0.052 |  |
|  | 17 | 5.5×10$^{-5}$ |  |  | -0.227 |  |
| $P_{z,2}$ | 130 | 3×10$^{-5}$ | 8.8×10$^{-4}$ | 0.18 | 0.033 | -0.08 |
|  | -20 | 1×10$^{-5}$ |  |  | -0.052 |  |
|  | 6 | 8.7×10$^{-5}$ |  |  | -0.227 |  |

Table. 1 Experimental values to obtain stiffness center **CR**$_{BT}$

The obtained results using this experimental approach are coherent with other finding in the literature [24].

Thereafter is made a verification measuring the tool point displacement O in work-space under the charge. We considerate our experimental results viewing the tool point that moves under one segment of a sphere with the center **CR**$_{BT}$: assimilated with a normal plan (**O, CR**$_{BT}$) for small displacements (accordingly figure 18).

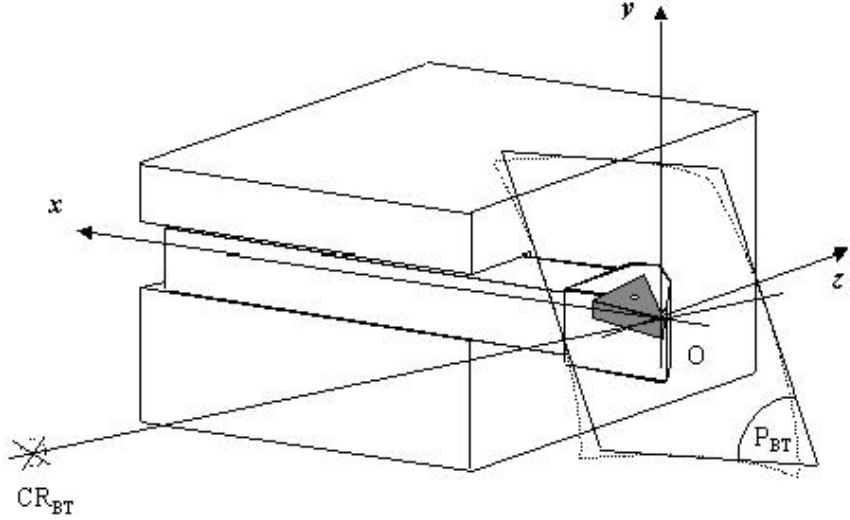

Fig. 18 Determination of the block tool center rotation: **CR**$_{BT}$

## *5.2 Comparaison with stiffness matrix*

If we proceed on the diagonalization of the block workpiece stiffness matrix [**K**$_{F,BT}$] obtained in the section (4.2.1}), we obtain:

$$[K_{F,\ BT\text{-}d}] = \begin{bmatrix} 4.1\times 10^5 & 0 & 0 \\ & 6\times 10^6 & 0 \\ & & 1.3\times 10^7 \end{bmatrix}_{O,xyz}. \quad (19)$$

The eigenvectors associated are:

$$[V] = \begin{bmatrix} v_1 & v_2 & v_3 \\ 0.0688 & 0.4103 & 0.6336 \\ 0.9896 & 0.3389 & -07713 \\ 0.1260 & 0.83467 & -0.0603 \end{bmatrix}_{O,xyz}. \quad (20)$$

We note that the maximum stiffness obtained is situated on the direction of third eigenvector $\vec{v_3}$, being at 4° nearly the direction (**CR**$_{BT}$ - O}) (see figure 19); and respectively the minimum and stiffness average on the directions of eigenvectors $\vec{v_1}$ and $\vec{v_2}$.

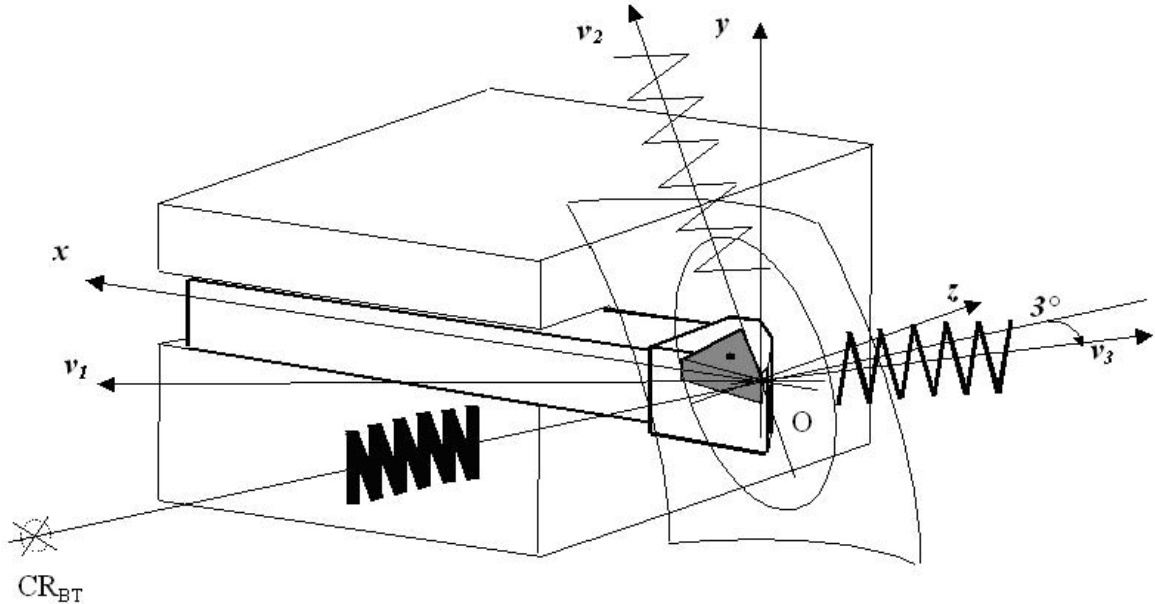

Fig 19 Eigenvectors.

## *6 Conclusion*

In the literature many authors use only tool rigidity [15, 31, 32, 40, 41] and recently the elastic behaviour of the machine [6]; furthermore, this research considers the rigidity of the tool device system (**BT**). Anothers takes into account that both the workpiece deflection and the cutting force between tool and workpiece [8]. In [34] tool and workpiece are modelled as two separate single degree of freedom spring-mass-damper systems. All these models are 2D, or more or less partially 3D, and thus does not allow a real 3D modeling of the cut.

The approach presented in this paper is fully different from previous studies in the sense that we use the six-dimension of the torsors (3 force components, 3 torque components). Consequently, this way allows us to be directly in the most suitable framework for a real 3D cutting modeling. Thanks to the principle of virtual work, we determined the complete torsor of small displacements (3 linear and 3 rotations displacement) associated with the mechanical actions torsor (3 force components, 3 torque components) via the matrix of rigidities. The stiffness center and the rotation center were obtained experimentally. The minimal displacement direction was defined on the basis of experimental model. Using Castigliano's theorem, we determined the angle $\alpha_K$ which characterizes the principal deformation direction. Then, we characterized block workpiece (**BW**) was characterized by its stiffness values and the diagonal matrix of displacements. This diagonal matrix of displacement validate the hypothesis and confirm the good rigidity of workpiece. Whole static behaviour analysis of the system (**BT-BW**) is thus validated. An application of this methodology within a framework of machining really 3D is available in [4]. These results are required to have a good 3D modeling of the cut, as being presented in a forthcoming paper.

## Acknowledgements

The authors acknowledge Jean Pierre Larivière, Ingineer CNRS (Centre National de la Recherche Scintifique - France) for the numerical simulation with SAMCEF software and Professor Miron Zapciu for the helpful discussions on this subject.

## Acknowledgements

The authors acknowledge Jean Pierre Larivière, Ingineer CNRS (Centre National de la Recherche Scintifique - France) for the numerical simulation with SAMCEF software and Professor Miron Zapciu for the helpful discussions on this subject.

## *References*

1. Axinte D. A., Belluco W., De Chiffre L.,
   Evaluation of cutting force uncertainty components in turning,
   International Journal of Machine Tools and Manufacture, **41**, pp. 719-730, (2001)

2. Benardos P. G., Mosialos S., Vosniakos G. C.,
   Prediction of workpiece elastic deflections under cutting forces in turning,
   Robotics and Computer-Integrated Manufacturing, **22**, pp. 505-514, (2006)

3. Bisu C. F.,
   Etude des vibrations auto-entretenues en coupe tridimensionnelle: nouvelle modélisation appliquée au tournage,
   Ph. D. Thesis, Université Bordeaux 1 and Universitatea Politehnica Bucharest, (2007)

4. Bisu C. F., Darnis P., Gérard A., K'nevez J-Y.,
   Displacements analysis of self-excited vibrations in turning,
   International Journal of Advanced Manufacturing Technology, **44**, 1-2, pp. 1-16, (2009)
   (doi: 10.1007/s00170-008-1815-8)

5. Buyuksagis I. S.,
   Analysis of circular marble sawing using a block-cutter,
   Ph. D. Thesis, Osmangazi University, Institute of Sciences and Technology, (1998)

6. Cano T., Chapelle F., Lavest J.-M., Ray P.,
   A new approach to identifying the elastic behaviour of a manufacturing machine,
   International Journal of Machine Tools and Manufacture, **48**, pp. 1569-1577, (2008)

7. Cardi A. A., Firpi H. A., Bement M. T., Liang S. Y.,
   Workpiece dynamic analysis and prediction during chatter of turning process,
   Mechanical Systems and Signal Processing, **22**, 1481-1494, (2008)

8. Carrino L., Giorleo G., Polini W., Prisco U.,
   Dimensional errors in longitudinal turning based on the unified generalized mechanics of cutting approach. Part I: Three-dimensional theory,
   International Journal of Machine Tools and Manufacture, **42**, pp. 1509-1515, (2002)

9. Castro L. R., Viéville P., Lipinski P.,
   Correction of dynamic effects on force measurements made with piezoelectric dynamometers,
   International Journal of Machine Tools and Manufacture, **46**, (14), pp. 1707-1715, (2006)

10. Chen C. K., Tsao Y. M.,
   A stability analysis of regenerative chatter in turning process without using tailstock,
   International Journal of Advanced Manufacturing Technology, **29**, (7-8), pp. 648-654, (2006)

11. Chen C. K., Tsao Y. M.,
   A stability analysis of turning tailstock supported flexible work-piece,
   International Journal of Machine Tools and Manufacture, **46**, (1), pp. 18-25, (2006)

12. Couétard Y.,
   Caractérisation et étalonnage des dynamomètres à six composantes pour torseur associé à un système de forces,
   Ph. D. Thesis, Universit\'e Bordeaux~1 Talence, (2000)

13. Dassanayake A. V., Suh C. S.,
   On nonlinear cutting response and tool chatter in turning operation,
   Communications in Nonlinear Science and Numercial Simulation, **13**, (5), pp. 979-1001, (2008)

14. Deacu I., Pavel G.,
   Vibrations des Machines-Outils,
   Dacia, Cluj Napoca, (1977)

15. Dimla Sr D. E.,
   The impact of cutting conditions on cutting forces and vibration signals in turning with plane face geometry inserts,
   Journal of Materials Processing Technology, **155-156**, pp. 1708-1715, (2004)

16. Ganguli A., Deraemaeker A., Preumont A.,
   Regenerative chatter reduction by active damping control,
   Journal of Sound and Vibration, **300**, pp. 847-862, (2007)

17. Gorodetskii Y. I., Budankov A. S., Komarov V. N.,
   A system for experimental studies of the dynamics of the process of cutting metal,
   Journal of Machinery Manufacture and Reliability, **37**, (1), pp. 68-73, (2008)

18. Insperger T., Barton D. A. W., Stepan G.,
   Criticality of Hopf bifucation in state-dependent delay model turning processes,
   International Journal of Non-Linear Mechanics, **43**, pp. 140-149, (2008)

19. Ispas C., Gheorghiu H., Parausanu I., Anghel V.,
   Vibrations des systèmes technologiques,
   Agir, Bucarest, (1999)

20. Karabay S.,
   Design criteria for electro-mechanical transducers and arrangement for measurement cutting forces acting on dynamometers,
   Materials & Design, **28**, pp. 496-506, (2007)

21. Koenigsberger F., Tlusty J.,
   Machine Tools Structures, Pergamon Press, (1970)

22. Konig W., Sepulveda E., Lauer-Schmaltz H.,
   Zweikomponenten schnittkraftmesser,
   Industrie-Anzeiger, (1997)

23. Korkut I.,
   Design and manufacturing of a dynamometer connected to computer which can do measuring with strain gages on the lathe,
   Ph. D. Thesis, University of Gazi, Institute of Science and Technology, (1996)

24. Kudinov V. A.,
   Dinamica Masinilor Unelten,
   Tehnicas, Bucarest,(1970)

25. Lapujoulade F., Coffignal G., Pimont, J.,
   Cutting forces evaluation during high speed milling,
   $2^{th}$ IDMME' 98, **2**, pp. 541-549,
   Compiègne, France, May, (1998)

26. Marinescu I., Ispas C., Boboc, D.,
   Handbook of Machine Tool Analysis,
   Deckker M., New York, (2002)

27. Mehdi K., Rigal J-F., Play D.,
   Dynamic behavior of thin wall cylindrical workpiece during the turning process, Part 1: Cutting process simulation,
   J. Manuf. Sci. and Engng., **124**, pp. 562-568, (2002)

28. Pérez H., Vizan A., Hernandez J. C., Guzman M.,
   Estimation of cutting forces in micromilling through the determination of specific cutting pressures,
   Journal of Materials Processing Technology, **190**, pp. 18-22, (2007)

29. Pestel E. C., Leckie F. A.,
   Matrix methods in elastomechanics,
   McGraw-Hill, New York, (1963)

30. Robinson J.,
   Analyse matricielle des structures à l'usage des ingénieurs,
   Dunod, Paris, (1971)

31. Saglam H., Unsacar F., Yaldiz S.,
   Investigation of the effect of rake angle and approaching angle on main cutting force and tool tip temperature,
   International Journal of Machine Tools and Manufacture, **46**, (2), pp. 132-141, (2006)

32. Saglam H., Yaldiz S., Unsacar F.,
   The effect of tool geometry and cutting speed on main cutting force and tool tip temperature,
   Materials & Design, **28**, pp. 355-360, (2002)

33. Salgado M. A., Lopez de Lacalle L.N., Lamikiz A., Munoa J., Sanchez J. A.,
   Evaluation of the stiffness chain on the deflection of end-mills under cutting forces,
   International Journal of Machine Tools and Manufacture, **45**, pp. 727-739, (2005)

34. Seka, M., Srinivas J., Kotaiah K. R., Yang S. H.,
   Stability analysis of turning process with tailstock-supported workpiece,
   International Journal of Advanced Manufacturing Technology, doi: 10.1007/s00170-0008-1764-2, (2008)

35. Seker U., Kurt A., Ciftci I.,
   Design and constrution of a dynamometer for measurement of cutting forces during machining with linear motion,
   Materials & Design, **23**, pp. 355-360, (2002)

36. Toh C. K.,
   Static and dynamic cutting force analysis when high speed rough milling hardened steel,
   Materials & Design, **25**, pp. 41-50, (2004)

37. Toulouse D.,
   Contribution à la modélisation et à la métrologie de la coupe dans le cas d'un usinage tridimensionnel, Ph. D. Thesis, Université Bordeaux 1 Talence, (1998)

38. Wang Z. C., Cleghorn W. L.,
   Stability analysis of spinning stepped-shaft workpieces in a turning process,
   Journal of Sound and Vibration, **250,** (2), pp. 356-367, (2002)

39. Yaldiz S., Ünsacar F.,
   Design, development and testing of a turning dynamometer for cutting force measurement,
   Materials & Design, **27**, 839-846, (2006)

40. Yaldiz S., Ünsacar F.,
   A dynamometer design for measurement the cutting forces on turning,
   Measurement, **39**, pp. 80-89, (2006)

41. Yaldiz S., Ünsacar F., Saglam H.,
   Comparaison of experimental results obtained by designed dynamometer to fuzzy model for predicting cutting forces in turning,
   Materials & Design, **27**, pp.1139-1147, (2006)